\documentclass[aps,pra,onecolumn,12pt]{revtex4-1}
\usepackage[dvipdfm]{graphicx}
\usepackage[usenames,dvipsnames]{xcolor}
\usepackage[dvipdfm]{hyperref}
\usepackage{amsmath}
\usepackage[makeroom]{cancel}
\begin{document}

\title{Generalized Hertz vector in the dissipative electrodynamics}
\author{Katalin Gamb\'ar}
	\email[Corresponding author: ]{gambar.katalin@uni-obuda.hu}
	\affiliation{Department of Natural Sciences, Institute of Electrophysics, \\ Kálmán Kandó Faculty of Electrical Engineering, \\ Óbuda University, \\ Tavaszmező u. 17, H-1084 Budapest, Hungary}
    
    \affiliation{Department of Natural Sciences, \\ National University of Public Service, \\  Ludovika t\'er 2, H-1083 Budapest, Hungary}
\author{Ferenc M\'arkus}
	\email[ ]{markus.ferenc@ttk.bme.hu; markus@phy.bme.hu}
	\affiliation{Department of Physics,
		Budapest University of Technology and Economics,
		H-1111 Budafoki \'ut 8., Budapest, Hungary}

\date{\today}

\begin{abstract}
In the electromagnetic theory, the Hertz vector reduces the number of potentials in the free fields. The further advantage of this potential is that it is much easier to solve some radiation processes. It indicates that the related method is sometimes more effective than the scalar and vector potential-based relations. Finally, the measurable field variables, the electric and magnetic fields, can be deduced by direct calculation from the Hertz vector. However, right now, the introduction of the Hertz vector operates if the conductive currents ${\bf j} = {\sigma}{\bf E}$ are neglected. We suggest a generalization for the case of conductive currents, i.e., for such cases when the electromagnetic field dissipates irreversibly into Joule heat. The presented procedure enables us to introduce also the Lagrangian formulation of the discussed dissipated electromagnetic waves. It opens a new way for future studies.
\end{abstract}

\maketitle



\section{Introduction: Reversible Electromagnetic Field}

\noindent Certain mathematical operators, such as the first derivatives, can bring irreversible and dissipative behavior in theory, as in the case of thermal conduction and damped oscillator \cite{mg1991,nyiri,gmny,gm1994,szegleti2020,markus_gambar_2022}. This minor change in the structure of equations causes a particular challenge in the Lagrangian construction \cite{rayleigh,mor,lebon1}. Historically it was a long time to find the mathematically consequent steps to the formulation method. Now, we will see that the Joule heat brings another twist to the theory. \\
In the case of free electrodynamic fields, the introduction of vector potential is necessary to deduce from Lagrangian formulation since the Maxwell equations contain non-selfadjoint operators (first-time derivative and divergence). The vector and scalar potentials generate the measurable physical variables. Mathematically these potentials are similarly applied, such as in the previous dissipative processes \cite{mg1991,gm1994,szegleti2020}. In general, we can recognize that the introduction of the potentials is independent of dissipation. (See a detailed comparative study about these requirements \cite{gllb2016}.) \\
We aim to achieve the Hamiltonian structure of dissipative electrodynamics. For this reason, we shortly summarize the concepts of the reversible case that can be considered classical theory. The complete system of the Maxwell equations \cite{jackson} - applying the usual notations – is
\begin{eqnarray}
rot {\bf H} &=& {\bf j} + \dot{\bf D}, \nonumber \\
rot {\bf E} &=& - \dot{\bf B}, \nonumber \\
div {\bf D} &=& {\varrho}, \nonumber \\ 
div {\bf B} &=& 0. 
\end{eqnarray}
If we restrict our examination for the free electromagnetic field, than the relation between the field quantities are ${\bf D} = {\varepsilon}_{0}{\bf E}$, ${\bf B} = {\mu}_{0}{\bf H}$, and both the current density and the charge density will be ${\bf j}=0$ and ${\varrho}=0$. The "dot" denotes the partial time derivative. The measurable field variables ${\bf E}$ and ${\bf B}$ can be expressed by introducing the vector potential ${\bf A}$ and the scalar potential $\varphi$
\begin{eqnarray}
{\bf E} &=& - \dot{\bf A} - grad {\varphi}, \nonumber \\
{\bf B} &=& rot {\bf A}.
\end{eqnarray}
It is worth mentioning, that originally the term $- \dot{\bf A}$ was introduced by Faraday as an electrostatic force relating to the so-called "electrotonic state" \cite{faraday1855,doncel1996}. Fortunately, the usage of the scalar and vector potential is also possible in the case of existence of charges and currents \cite{konopinski1978,gambar_rocca_markus_2020}. Now, we can write the Lagrange density function $L$ 
\begin{equation}
L = \frac{1}{2} {\varepsilon}_{0} 
\left( - \dot{\bf A} - grad {\varphi} \right)^{2} -
\frac{1}{2{\mu}_{0}} \left( rot {\bf A} \right)^{2} 
\end{equation}
We calculate the equations of motion of the problem varying concerning ${\bf A}$ and $\varphi$
\begin{equation}
{\varepsilon}_{0}{\mu}_{0} \ddot{\bf A} - {\Delta}{\bf A} = 0  \label{wave_eq_A}
\end{equation}
and
\begin{equation}
{\varepsilon}_{0} div(-\dot{\bf A} - grad{\varphi}) =
div{\bf E} = 0,
\end{equation}
where we have to take into account the Lorentz condition
\begin{equation}
div{\bf A} + {\varepsilon}_{0}{\mu}_{0} \dot{\varphi} = 0.
\end{equation}
We can understand from Eq. (\ref{wave_eq_A}) why the free field equations with the first order derivatives do not mean an irreversible process. The derived Eq. (\ref{wave_eq_A}) is a non-dissipative wave equation for free propagation. So, in general, a careful examination is needed before the Lagrangian elaboration of different theories.  
Instead of the two potentials, Hertz introduced a vector ${\bf \Pi}$ with the definitions
\begin{eqnarray}
{\bf A} &=& {\varepsilon}_{0}{\mu}_{0} \dot{\bf \Pi}, \nonumber \\
{\varphi} &=& - div{\bf \Pi}
\end{eqnarray}
\noindent to apply only one vector (potential) field. The fields ${\bf E}$ and ${\bf B}$ are by the Hertz vector
\begin{eqnarray}  \label{Hertz}
{\bf E} &=& - {\varepsilon}_{0}{\mu}_{0} \ddot{\bf \Pi} +
grad \, div{\bf \Pi}, \nonumber \\
{\bf B} &=& {\varepsilon}_{0}{\mu}_{0} 
rot \dot{\bf \Pi}.
\end{eqnarray}
This fact shows that we can deduce the free electromagnetic field from one generator field \cite{jackson,stratton}. There are electromagnetic problems that cannot be solved or are complicated to solve without the Hertz vector \cite{essex1977}. The symmetry of the Maxwell equations allow an alternative introduction of the Hertz vector involving its gauge symmetries \cite{kannanberg1987,gough1996,elbistan2018}. Ornigotti et al. have shown that due to the transversality of the electromagnetic wave, the Hertz vector can be expressed as a product of a constant polarization vector and a scalar potential \cite{ornigotti2014}. It may give a next physical conception to the Hertz vector formulations.

\section{Generalized vector potential, Hertz vector and modified Lorenz condition}

In the presence of electrically conductive materials, the currents influence the electromagnetic field. So,
we take into account the conductive current density by the differential Ohm’s law ${\bf j} = {\sigma}{\bf E}$, thus the
Maxwell equations are
\begin{eqnarray}  \label{Maxwell-eqs}
\frac{1}{{\mu}_{0}}rot {\bf B} &=& {\sigma}{\bf E} + {\varepsilon}_{0}\dot{\bf E}, \nonumber \\
rot {\bf E} &=& - \dot{\bf B}, \nonumber \\
div {\bf E} &=& 0, \nonumber \\ 
div {\bf B} &=& 0. 
\end{eqnarray}
From a thermodynamical viewpoint, its importance of it is unquestionable. The difficult question is how to generalize the above formalism for the present case \cite{stratton}. It can be seen that the new term ${\sigma}{\bf E}$ causes difficulties. Let us introduce a generalized definition of the vector potential ${\bf A}_{m}$ and the Hertz vector ${\bf \Pi}_{m}$ involving the current related term 
\begin{eqnarray}  \label{generalized_vector_potential}
{\bf A}_{m} &=& {\varepsilon}_{0}{\mu}_{0} \dot{\bf \Pi}_{m} + {\sigma}{\mu}_{0}{\bf \Pi}_{m}, \nonumber \\
{\varphi} &=& - div{\bf \Pi}_{m},
\end{eqnarray}
and the modified Lorenz condition 
\begin{equation}
div{\bf A}_{m} + {\varepsilon}_{0}{\mu}_{0} \dot{\varphi} + {\sigma}{\mu}_{0}{\varphi} = 0.
\end{equation}
Here, we can point out the role of Hertz vector ${\bf \Pi}_{m}$. We formulate the field quantities ${\bf E}$ and ${\bf B}$ are similar to Eqs. (\ref{Hertz}), but now we add a new term in each equation
\begin{eqnarray}  \label{genHertz}
{\bf E} &=& - {\varepsilon}_{0}{\mu}_{0} \ddot{\bf \Pi}_{m} -
{\sigma}{\mu}_{0} \dot{\bf \Pi}_{m} + grad \, div{\bf \Pi}_{m}, \nonumber \\
{\bf B} &=& {\varepsilon}_{0}{\mu}_{0} rot \dot{\bf \Pi}_{m} + {\sigma}{\mu}_{0} rot {\bf \Pi}_{m}.
\end{eqnarray}
\noindent We can conclude that a single generator space is still sufficient to produce gauge spaces. The connections between the measurable field quantities ${\bf E}$, ${\bf B}$ and the potentials ${\bf A}_{m}$ and $\varphi$ remain the same physical meaning
\begin{eqnarray}
{\bf E} &=& - \dot{\bf A}_{m} - grad {\varphi}, \nonumber \\
{\bf B} &=& rot {\bf A}_{m}.
\end{eqnarray}
One can prove that all of the field quantities complete the requirement of damping wave (telegrapher) equation
\begin{equation}
0 = {\varepsilon}_{0}{\mu}_{0} \ddot{\bf G} + {\sigma}{\mu}_{0} \dot{\bf G} - \triangle {\bf G},
\end{equation}
where ${\bf G}$ can be ${\bf E}$, ${\bf B}$, ${\bf A}_{m}$, $\varphi$ and ${\bf \Pi}_{m}$. By the first equation of Eq. (\ref{genHertz}), the electronic field ${\bf E}$ can be expressed in an alternative form
\begin{equation}
{\bf E} = rot rot {\bf \Pi}_{m}.
\end{equation}

\section{The Lagrangian formulation}

\noindent We need to find a suitable Lagrange density function if we would like to deduce the field equation and exploit the Hamiltonian principle.   The construction is not self-explanatory. However, if it is possible, the existence of Lagrangian is of fundamental importance. Now, the formulated Lagrangian is
\[
L = \frac{1}{2} {\varepsilon}_{0} \left( - \dot{\bf A}_{m} - grad {\varphi} \right)^2 - 
\frac{1}{2{\mu}_{0}} \left( rot {\bf A}_{m} \right)^2 + {\sigma} rot {\bf \Pi}_{m} rot {\bf A}_{m} 
\]
\begin{equation}  \label{E-M_Lagrangian}
-\frac{1}{2} {\sigma}^2 {\mu}_{0} \left( rot {\bf \Pi}_{m} \right)^2 
-\frac{1}{2} {\varepsilon}_{0} {\bf \Pi}_{m} rot rot {\Delta} {\bf \Pi}_{m} 
-\frac{1}{2} {\varepsilon}_{0}^2 {\mu}_{0} \left( rot \dot{{\bf \Pi}}_{m} \right)^2 ,
\end{equation}
\noindent which pertains to a dissipative process in electrodynamics. The eleboration of variation is necessary for each field function as variables ${\bf A}_{m}$, ${\varphi}$ and ${\bf \Pi}_{m}$. The exactness of Lagrangian is complete if we obtain the relevant equations of motion or field equations. For the easy following, the details of the calculation are also displayed. \\
{\it a}, With respect to ${\bf A}_{m}$: 
\[
0 = -\frac{\partial}{{\partial}t} \frac{{\partial}L}{{\partial}\dot{\bf A}_{m}} + rot \frac{{\partial}L}{{\partial}rot{\bf A}_{m}} 
\]
\[
= {\varepsilon}_{0} \frac{\partial}{{\partial}t} \left( - \dot{\bf A}_{m} - grad {\varphi} \right) - \frac{1}{{\mu}_{0}} rot rot {\bf A}_{m} + {\sigma} rot rot {\bf \Pi}_{m}
\]
\begin{equation}
= {\varepsilon}_{0} \dot{\bf E} - \frac{1}{{\mu}_{0}} rot {\bf B} + {\sigma} {\bf E}.
\end{equation} 
This is exactly the first equation in the Maxwell equations in Eqs. (\ref{Maxwell-eqs}). \\
{\it b}, With respect to $\varphi$:
\begin{equation}
0 = -div \frac{{\partial}L}{{\partial} grad{\varphi}} = {\varepsilon}_{0} div \left( - \dot{\bf A}_{m} - grad {\varphi} \right) = {\varepsilon}_{0} div {\bf E},
\end{equation}

which is the third equation in Eqs. (\ref{Maxwell-eqs}). \\
{\it c}, With respect to ${\bf \Pi}_{m}$:
\[
0 = \frac{{\partial}L}{{\partial} {\bf \Pi}_{m}} + rot \frac{{\partial}L}{{\partial} rot {\bf \Pi}_{m}} +
rot rot {\Delta} \frac{{\partial}L}{{\partial} rot rot {\Delta} {\bf \Pi}_{m}} 
- rot \frac{\partial}{{\partial} t} \frac{{\partial}L}{{\partial} rot \dot{\bf \Pi}}_{m}
\]
\[
= -\frac{1}{2} {\varepsilon}_{0} rot rot {\Delta} {\bf \Pi}_{m} + {\sigma} rot rot {\bf A}_{m} - {\sigma}^2 {\mu}_{0} rot rot {\bf \Pi}_{m} - \frac{1}{2} {\varepsilon}_{0} rot rot {\Delta} {\bf \Pi}_{m} + {\varepsilon}_{0}^2 {\mu}_{0} rot rot \ddot{{\bf \Pi}}_{m}
\]
\begin{equation}
= - {\varepsilon}_{0} rot rot {\Delta} {\bf \Pi}_{m} + {\sigma} rot rot {\bf A}_{m} - {\sigma}^2 {\mu}_{0} rot rot {\bf \Pi}_{m} + {\varepsilon}_{0}^2 {\mu}_{0} rot rot \ddot{{\bf \Pi}}_{m}
\end{equation}
Applying the definition of generalized vector potential ${\bf A}_{m}$ in Eq. (\ref{generalized_vector_potential}), we obtain
\[
0 = - {\varepsilon}_{0} rot rot {\Delta} {\bf \Pi}_{m} + \sigma {\varepsilon}_{0}{\mu}_{0} rot rot \frac{\partial{\bf \Pi}_{m}}{{\partial}t} + \cancel{ {\sigma}^2 {\mu}_{0} rot rot {\bf \Pi}_{m} } - \cancel{ {\sigma}^2 {\mu}_{0} rot rot {\bf \Pi}_{m} } + {\varepsilon}_{0}^2 {\mu}_{0} rot rot \ddot{{\bf \Pi}}_{m}
\]
After the above simplification, dropping the rotrot operator and dividing by ${\varepsilon}_{0}$, the telegrapher's equation appears for the generalized Hertz vector ${\bf \Pi}_{m}$:
\begin{equation}
0 = {\varepsilon}_{0} {\mu}_{0} \ddot{{\bf \Pi}}_{m} + \sigma {\mu}_{0} \dot{\bf \Pi}_{m} - {\Delta} {\bf \Pi}_{m},
\end{equation}
that is the expected telegrapher equation. We can conclude that the Lagrangian in Eq. (\ref{E-M_Lagrangian}) involves and describes well the Joule dissipation. Now, the irreversible behavior of the electromagnetic theory is within the frame of the Hamilton's principle.

\section{The electromagnetic energy loss}

The canonical momenta pertain to the field quantities that exist as pure time derivatives in the Lagrangian. Thus, we can define the canonical momentum just for ${\bf A}_{m}$ as it is usual in the Hamiltonian formulation
\begin{equation}  
P_{{\bf A}_{m}} = \frac{\partial L}{\partial \dot{\bf A}_{m}} = - {\varepsilon}_{0} \left( - \dot{\bf A}_{m} - grad {\varphi} \right).
\end{equation}
Thus the Hamiltonian of the dissipative electromagnetic field is
\[  
H = P_{{\bf A}_{m}} \dot{\bf A}_{m} - L 
\]

\[
= \underbrace{- {\varepsilon}_{0} \left( - \dot{\bf A}_{m} - grad {\varphi} \right) \dot{\bf A}_{m} - \frac{1}{2} {\varepsilon}_{0} \left( - \dot{\bf A}_{m} - grad {\varphi} \right)^2}_{\frac{1}{2} {\varepsilon}_{0} {\bf E}^2 + \underbrace{{\varepsilon}_{0} {\bf E} grad {\varphi}}_{1}} + \underbrace{\frac{1}{2{\mu}_{0}} \left( rot {\bf A}_{m} \right)^2}_{\frac{1}{2{\mu}_{0}}{\bf B}^2} - {\underbrace{{\sigma} rot {\bf \Pi}_{m} \cdot rot {\bf A}_{m}}_{{\sigma} rot {\bf \Pi}_{m} \cdot {\bf B}}}.
\]

\begin{equation}  \label{E-M_Hamiltonian_1_form}
+\frac{1}{2} {\sigma}^2 {\mu}_{0} \left( rot {\bf \Pi}_{m} \right)^2 
+ \underbrace{\frac{1}{2} {\varepsilon}_{0} {\bf \Pi}_{m} rot rot {\Delta} {\bf \Pi}_{m}}_{2} 
+ \underbrace{\frac{1}{2} {\varepsilon}_{0}^2 {\mu}_{0} \left( rot \dot{{\bf \Pi}}_{m} \right)^2}_{3} ,
\end{equation}
Since the Hamiltonian does not depend on time explicitly, thus volume integral of it is constant, i.e., the energy conservation is valid. \\
We obtain from the underbraced term 1:
\begin{equation}
{\varepsilon}_{0} {\bf E} grad {\varphi} = {\varepsilon}_{0} div ( {\bf E} {\varphi} ) - {\varepsilon}_{0} {\varphi} \underbrace{div {\bf E}}_{=0}. 
\end{equation}
The Hamiltonian is invariant against the divergence and time derivate terms. So, here, the term ${\varepsilon}_{0} div ( {\bf E} {\varphi} )$ can be dropped from the Hamiltonian density function. We can reformulate the underbraced terms 2 and 3 as
\begin{equation}
\frac{1}{2} {\varepsilon}_{0} {\bf \Pi}_{m} rot rot {\Delta} {\bf \Pi}_{m} = \frac{1}{2} {\varepsilon}_{0} {\bf \Pi}_{m} rot {\bf \dot{B}},
\end{equation}
\begin{equation}
\frac{1}{2} {\varepsilon}_{0}^2 {\mu}_{0} \left( rot \dot{{\bf \Pi}}_{m} \right)^2 = \frac{1}{2} {\varepsilon}_{0} rot \dot{{\bf \Pi}}_{m} \left( {\bf B} - \sigma {\mu}_{0} rot {\bf \Pi}_{m} \right).
\end{equation}
Thus, the Hamiltonian has an almost final form:
\[
H = \frac{1}{2} {\varepsilon}_{0} {\bf E}^2 + \frac{1}{2{\mu}_{0}}{\bf B}^2 + \frac{1}{2} \sigma^2 {\mu}_{0} \left( rot {\bf \Pi}_{m} \right)^2 \underbrace{- {{\sigma} rot {\bf \Pi}_{m} \cdot {\bf B}}}_{4} 
\]
\begin{equation}
+ \underbrace{\frac{1}{2} {\varepsilon}_{0} {\bf \Pi}_{m} rot {\bf \dot{B}} + \frac{1}{2} {\varepsilon}_{0} rot \dot{{\bf \Pi}}_{m} \left( {\bf B} - \sigma {\mu}_{0} rot {\bf \Pi}_{m} \right)}_{5}
\end{equation}
We need to discuss and reformulate these underbraced expressions. By the substitution of ${\bf B}$ from Eq. (\ref{genHertz}), the underbraced term 4 is
\begin{equation}
- {{\sigma} rot {\bf \Pi}_{m} \cdot {\bf B}} = - \frac{1}{2} {\varepsilon}_{0}{\mu}_{0} {\sigma} \frac{\partial}{\partial t} (rot {\bf \Pi}_{m})^2 - {\sigma}^2 {\mu}_{0} (rot {\bf \Pi}_{m})^2.
\end{equation}
The underbraced 5 is: 
\begin{equation}
\frac{1}{2} {\varepsilon}_{0} \frac{\partial}{\partial t} \left( {\bf \Pi}_{m} rot {\bf B} \right) \underbrace{- \frac{1}{2} {\varepsilon}_{0} \dot{{\bf \Pi}}_{m} rot {\bf B} + \frac{1}{2} {\varepsilon}_{0} rot \dot{{\bf \Pi}}_{m} \cdot {\bf B}}_{\frac{1}{2} {\varepsilon}_{0} div (\dot{{\bf \Pi}}_{m} \times {\bf B})} - \frac{1}{4} {\varepsilon}_{0} {\mu}_{0} \sigma \frac{\partial}{\partial t} \left( rot {\bf \Pi}_{m} \right)^2. 
\end{equation}
Since the total time derivatives and the divergences can be dropped from the Hamiltonian thus the 
\begin{equation}
H = \frac{1}{2} {\varepsilon}_{0} {\bf E}^2 + \frac{1}{2{\mu}_{0}}{\bf B}^2 - \frac{1}{2} \sigma^2 {\mu}_{0} \left( rot {\bf \Pi}_{m} \right)^2  
\end{equation}
can be deduced. Here, the third term pertains to the dissipative Joule heat loss caused by the conductive current. If an electric conductor does not take place in space, then the electromagnetic energy remains constant.

\section{Summary}

We pointed out that the Hertz vector can have a generalized form by which the Maxwell equations, involving the conductive currents, can be successfully produced. In this way, the Joule dissipation appears on a potential level. This generalization of the Hertz vector enables us to create the Lagrangian description of such an electromagnetic field in which we can handle the loss of electromagnetic energy. The calculated Hamiltonian of the process clearly shows that the electromagnetic field energy dissipates into Joule heat. If there is no conductive current in the space, the electromagnetic energy is conserved during the process. We hope that based on the presented Lagrangian formulation, the electromagnetic and the thermal fields can couple, by which further studies may be possible in the case of electromagnetic radiation in media. \\

{\bf Acknowledgment} \\
TKP2021-NVA-16 has been implemented with the support provided by the Ministry of Innovation and Technology of Hungary from the National Research, Development and Innovation Fund. This research was supported by the National Research, Development and Innovation Office (NKFIH) Grant Nr. K137852 and by the Ministry of Innovation and Technology and the NKFIH within the Quantum Information National Laboratory of Hungary.

\end{document}